\documentclass[%
twocolumn,
superscriptaddress,
amsmath,
amssymb, 
aps,
prl,
floatfix,
]{revtex4-2}

\usepackage{graphicx}
\usepackage{bm}
\usepackage{amsmath}
\usepackage[dvipsnames]{xcolor}
\usepackage{hyperref}
\usepackage[normalem]{ulem}
\newcommand{\threejm}[6]{ \left(\begin{array}{ccc} #1 & #3 & #5\\
                                              #2 & #4 & #6
                                \end{array}
                          \right)}
\newcommand{\sixj}[6]{ \left\{\begin{array}{ccc} #1 & #2 & #3\\
                                              #4 & #5 & #6
                                \end{array}
                          \right\}}

\begin{document}
\title{
Low-entropy arrays of microwave-shielded molecules prepared by interaction blockade
}

\author{Tijs Karman}
\email{t.karman@science.ru.nl}
\affiliation{Institute for Molecules and Materials, Radboud University, 6525 AJ Nijmegen, The Netherlands}
\author{Sebastian Will}
\affiliation{Department of Physics, Columbia University, New York, New York 10027, USA}
\author{Zoe Yan}
\email{zzyan@uchicago.edu}
\affiliation{James Franck Institute and Department of Physics, The University of Chicago, Chicago, IL 60637, USA}

\date{\today}

\begin{abstract}
Ultracold molecules are becoming an increasingly important technology for quantum simulation, computation, and sensing, but their state preparation in large, low-entropy arrays remains a key challenge.
We propose to deterministically load single molecules into optical tweezer arrays or lattices from either thermal or degenerate gases, with a high probability of occupying the tweezer's motional ground state. 
Strong repulsion between microwave-shielded molecules prevents multiparticle occupancy.
Our proposal represents a robust scheme for deterministic single molecule preparation directly in the motional ground state with expected fidelities exceeding 99\% for small trap volumes and highly polar species.
This method can be scaled to thousands of traps limited by the reservoir molecule number, opening the door to large, low-entropy polar molecule arrays for quantum computation, quantum simulation, and precision measurement.
\end{abstract}

\maketitle

Defect-free arrays of atoms and molecules form a powerful tool for modern quantum simulation, computing, and metrology~\cite{kaufman2021quantum}, allowing unparalleled control and detection capabilities over individual quantum particles.
Currently with atomic systems, as many as $\sim$6000 qubits can be prepared in large-scale tweezer arrays~\cite{manetsch2025tweezer}, and several methods exist to remove defects in occupancy via rearrangement~\cite{barredo2016atom, endres2016atom} and to remove entropy from thermal excitations by ground-state cooling of the atomic motion~\cite{Kaufman2012}.
Tremendous advances have also been made in generating tweezer arrays of directly laser-cooled~\cite{anderegg2019optical, holland2023demand} and assembled molecules~\cite{Cairncross2021,Ruttley2024}, overcoming challenges associated with their complex internal structure~\cite{cornish2024quantum, langen2024quantum}.
Low-entropy, large-scale tweezer arrays of molecules would present multiple opportunities across the spectrum of quantum sciences, and their creation is an outstanding challenge.  
For example, in quantum simulation, ultracold polar molecules have been proposed as a powerful platform for realizing lattice spin Hamiltonians relevant to quantum magnetism~\cite{Barnett2006, micheli2006toolbox, gorshkov2011}, but state-preparation challenges have so far limited molecular quantum simulators to studying disordered spin models~\cite{yan2013observation, christakis2023probing, carroll2025observation}.
Universal quantum computers based on molecular qubits have been proposed~\cite{DeMille2002}, 
and long-lived coherence \cite{ruttley2025long}
and 
entanglement have been demonstrated with individually controlled molecules~\cite{bao2023dipolar, holland2023demand, picard2025entanglement}, but high-fidelity state preparation remains a challenge. 
Finally, molecules are leading probes for precision measurements of parity and time-reversal
violating interactions
~\cite{Safronova2018}, where next-generation experiments may benefit from large arrays of trapped molecules~\cite{alarcon2022electricdipolemomentssearch}.

Broadly, there are two mechanisms for the preparation of particles in a conservative optical potential with sub-Poissonian statistics, loaded from a cold reservoir gas: (1) 
laser-cooling, in which a photon takes away excess energy, 
or (2)  collisions, in which a second particle removes the excess energy. 
Direct laser-cooling of molecules into optical tweezers has been demonstrated~\cite{anderegg2019optical, holland2023demand, bao2023dipolar};
the second molecule is ejected by light-assisted collisions, leading to zero or one particle with near equal probabilities,
and schemes for improved loading have been proposed~\cite{shaw2023dark,walraven2024scheme}.
Currently, rearrangement has created defect-free arrays of $\sim$10 sites~\cite{holland2023demand, bao2023dipolar}, and progress toward high-fidelity cooling of the molecular motion to the ground state is ongoing~\cite{bao2024raman, lu2024raman}. 
However, not all species can be laser-cooled, and the power budget requirement that allows laser-cooling in tweezers is quite high, potentially imposing a limit on the array size. 
With collision-based loading, a conceptually simple way to realize a perfectly-filled low-entropy molecular array is to load from a  quantum-degenerate bulk gas and to prepare a band/Mott insulator of fermions/bosons~\cite{Bloch2008}, though this has not been demonstrated to date.
Currently, polar molecules have been loaded into optical lattices from near-degenerate bulk gases or associated in lattices, resulting in occupancies below 30\%~\cite{yan2013observation,moses2015creation, Reichsollner2017, christakis2023probing, carroll2025observation}, or created out of pre-cooled, tweezer-trapped atoms that are then assembled via magnetic Feshbach resonances~\cite{Cairncross2021,Ruttley2024}, with state preparation currently at $\sim50$\%~\cite{Ruttley2024}.

\begin{figure}[t]
    \centering
    \includegraphics[width=0.475\textwidth]{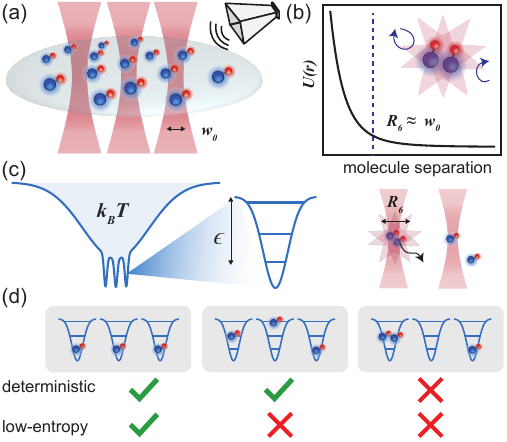}
    
\caption{\label{fig:1} {\bf Illustration of deterministic loading of single molecules via interaction blockade}. 
\textbf{(a)} A red optical tweezer with waist $w_0$ is overlapped with thermal reservoir gas (blue) with many polar molecules, which experience repulsion from microwave shielding, \textbf{(b)} with intermolecular energy shown versus distance.  The length scale for the $c_6$ potential, $R_6$, can approach or exceed the tweezer waist, blocking a second molecule from remaining inside the tweezer.
\textbf{(c)}
Energy scales relevant to our model include the reservoir thermal energy $k_BT$ and the single-particle binding energy $\epsilon$ of the tweezer.
The deterministic single-particle loading pictured in (a) will occur approximately in the conditions that $U\gg\epsilon\gg k_BT$.
 As molecules collide inside a small tweezer volume, a sufficiently strong two-particle repulsive energy $U$ guarantees that multiparticle occupancy is not supported in thermal equilibrium. 
 \textbf{(d)} Compared to stochastic loading from a thermal gas, which produces disordered arrays (right), our scheme can deterministically load single molecules in tweezer traps (middle) with high probability of ground state occupancy (left), leading to low-entropy arrays.
}
\end{figure}

Here we propose using collisions from repulsive interactions to deterministically load single molecules into a tight optical tweezer or lattice potential.
The idea is illustrated in Fig.~\ref{fig:1}.
Collisions load molecules from a bulk gas into the tweezer, which remains in thermal contact.
Strong intermolecular repulsion prohibits multi-particle occupancy of the tweezer, assuming thermal equilibrium.
Conceptually, an interaction blockade forms when the repulsion length scale approaches or exceeds the spatial extent of the tweezer volume (set by the waist size $w_0$ and wavelength $\lambda$).
We show that not only can the resulting tweezer be loaded with near-unit probability, but the particle also preferentially occupies the motional ground state of the trap, leading to \textit{low-entropy} tweezer arrays. 
The strong intermolecular repulsion is engineered with the versatile \textit{microwave shielding} technique~\cite{karman2018microwave, Lassabliere2018}, which simultaneously suppresses two-body inelastic collisions in the gas.
Practical implementation likely involves dynamical ramping of the tweezer potential and control of molecular interactions,
which may lead to further enhancements, but we consider equilibrium populations as a conservative bound applicable to sufficiently slow ramps.

Molecules undergo collisional loss and shielding is necessary for collisional stability of the reservoir \cite{bigagli2023collisionally}.
Collisional shielding 
spectacularly has enabled evaporative cooling to quantum degenerate gases of fermionic~\cite{valtolina2020dipolar,li2021tuning,schindewolf2022evaporation} and bosonic molecules~\cite{bigagli2024observation, shi2025bose}.
There exist various schemes for shielding \cite{gorshkov2008suppression,gonzalez2017adimensional,augustovivcova2019ultracold,karam2023two,karman2018microwave,Lassabliere2018,karman2025double}.
We will focus on double microwave shielding with compensated dipolar interactions \cite{karman2025double}.
The interaction between shielded molecules is given by a repulsive van der Waals shield, $+c_6 r^{-6}$, pictured in Fig.\ref{fig:1}(b), where $r$ is the intermolecule separation and the $c_6$ coefficient is molecule- and microwave-dependent.
The strength of the repulsion is characterized by a length scale $R_6 = (\frac{m c_6}{\hbar^2})^{1/4}$, where $m$ is the mass.
In current experiments, $R_6$ is around 160~nm for sodium-cesium (NaCs) \cite{bigagli2023collisionally,bigagli2024observation}, a bialkali molecule with currently the highest electric dipole moment realized in the laboratory of $d_0 = 4.6$ Debye (D)~\cite{Cairncross2021, Stevenson2023}.
This length scale may even be substantially enhanced for more polar molecules,
estimated using the universality of microwave shielding \cite{dutta2025universality} as $R_6 \approx 480$~nm for potassium-silver (KAg, 8.5\,D~\cite{Smialkowski}) and $R_6 \approx 760$~nm for francium-silver (FrAg, 9.2\,D~\cite{Smialkowski}).
The latter two belong to a class of ``ultrapolar" molecules of assembled alkali-metal-coinage-metal atoms~\cite{Smialkowski}, which are expected to produce among the highest achievable dipole moments in diatomic molecules 
due the coinage metal’s electron affinity, and are useful in next-generation precision measurements of fundamental physics~\cite{sunaga2019merits, fleig2021theoretical, klos2022prospects, marc2023candidate, polet2024p, marc2025semi}.
There are experimental efforts underway for the alkali-coinage metal species~\cite{Vayninger2025, demilleLab}.

Especially for the ultrapolar molecules, $R_6$ can exceed the tweezer waist, $w_0$,
and one may envision the shield blocking enough of the tweezer volume to prevent the existence of any two-molecule tweezer-bound states.
However, the shielding potential is not a hard wall,
but rather the interaction energy at a distance $R_6$ is on the order of $E_6 = \hbar^2/2mR_6^2$ which is below $h\times1$kHz,
implying that even if $R_6>w_0$, two-molecule bound states are fully expelled only for extremely low tweezer depths
that may present challenges for practical utilization.
However, we will show that this condition of completely expelling two-molecule bound states will not be necessary for low-entropy tweezer loading; multiparticle occupancies can be made extremely unlikely when the repulsion energy exceeds the single-particle binding energy.
We present calculations of our scheme for NaCs, KAg, and FrAg.

We model single-molecule loading by considering a single optical tweezer equilibrated in a bath of molecules.
Specifically, we consider a reservoir of $N$ molecules in a box trap of volume $V$ and temperature $T$ -- though the precise trapping potential is not essential.
The system has total energy $E_\mathrm{tot} = \frac{3}{2} N k_BT$.
The reservoir is brought in thermal contact with a tweezer potential that supports a single-particle bound state with binding energy $\epsilon>0$ below the potential energy in the box trap.
We determine the equilibrium occupation by considering the statistical weights of configurations with various numbers of particles loaded into the tweezer.
The statistical weight for an unoccupied tweezer is proportional to the number of microstates
available to the reservoir
\begin{align}
W(N,E_\mathrm{tot},V) = \frac{V^N}{N!~h^{3N}}\frac{\pi^{3N/2}~2}{\Gamma(3N/2)} (2mE)^{\frac{3N-1}{2}},
\end{align}
where $\Gamma$ is the gamma function.
If a single molecule is loaded into the tweezer,
the number of microstates accessible to the reservoir is $W(N-1,E_\mathrm{tot}+\epsilon,V)$, due particle number and energy conservation.
Hence, in the thermodynamic limit, the probability of the tweezer hosting exactly one particle, $p_1$, versus zero, $p_0$, is given by
\begin{align}
\frac{p_1}{p_0} = \frac{W(N-1,E_\mathrm{tot}+\epsilon,V)}{W(N,E_\mathrm{tot},V)} \simeq \frac{N\Lambda^3}{V} \exp\left( \frac{\epsilon}{k_BT} \right),
\label{eq:p1p0}
\end{align}
where $\Lambda=h/\sqrt{2\pi m k_B T}$ is the thermal de Broglie wavelength.
The pre-exponential factor $\rho\equiv N\Lambda^3/V$ is the phase-space density.
Next we consider configurations corresponding to two molecules loaded into the same tweezer, bound by a combined energy of $2\epsilon-U$, with $U>0$ the repulsive interaction energy between the two tweezer-trapped molecules.
This configuration has weight $\frac{1}{2!} W(N-2,E_\mathrm{tot}+2\epsilon-U,V)$
and
for the relative occupation we find
\begin{align}
\frac{p_2}{p_1} &= \frac{1}{2!} \rho \exp\left( \frac{\epsilon-U}{k_BT} \right).
\label{eq:p2p1}
\end{align}

The same occupations can be found using the grand canonical ensemble,
where for the classical gas the fugacity $z\equiv \exp(\mu/k_BT) = \rho$, where $\mu$ is the chemical potential.
For a degenerate Bose gas, the fugacity is instead determined by $\mathrm{Li}_{3/2}(z) = \rho$, where $\mathrm{Li}_{3/2}$ is the poly-logarithmic function,
and the factor $2!$ disappears as the two-molecule states are restricted by exchange symmetry \cite{huang1963statistical}.

\begin{figure}[htb]
    \centering
    \includegraphics[width=0.47\textwidth]{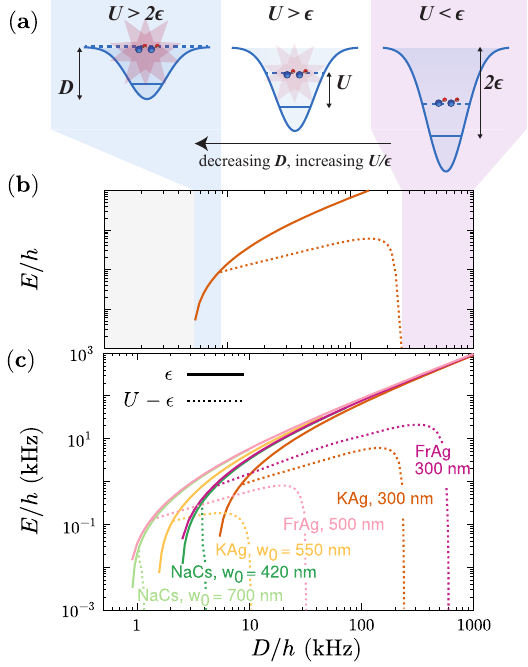}
\caption{
{\bf Blocking double occupancy by the interaction blockade}.
Panel ({\bf a}) illustrates interacting (non-interacting) two-molecule energy levels as dashed (solid) levels in tweezers of varying depth $D$.
In deep tweezers [red shading], the repulsion $U$ is small compared to the binding energy,
and double occupancy is not prevented.
At intermediate depths [white shading], two-molecule bound states exist but are energetically unfavored for $U>\epsilon$,
and only at the shallowest depths [blue shading], two-molecule bound states are fully expelled.
Panel ({\bf b}) illustrates the single particle binding energy $\epsilon$ and energetic cost to loading a second molecule, $U-\epsilon$, for KAg in a 300\,nm tweezer.
Results of DVR calculations for other species and tweezer parameters are shown in panel ({\bf c}).
\label{fig:dvr} }
\end{figure}

Repulsive interactions between the molecules can lead to the desired sub-Poissonian statistics.
In particular, the probability of loading a second molecule can be made exponentially small at low temperature as long as $U>\epsilon$,
while the probability of a tweezer remaining empty can simultaneously be exponentially small for $\epsilon > k_B T$.
The criterion $U>\epsilon$ essentially means that there is an energetic cost to loading a second molecule,
and this is unlikely to occur in equilibrium because it reduces entropy in the reservoir.
We note that this criterion is far easier to fulfill than to expel all two-body bound states, which requires $U> 2\epsilon$, and we argued above that collisional shielding may not realize this even when the $R_6$ range of the repulsive shield exceeds the tweezer waist.
In the Appendix, we provide a conservative estimate that higher occupation is also suppressed in this regime.
For simplicity we have discussed the case where the tweezer supports a single state,
but this simplification is not essential and the Appendix discusses loading a single molecule in the presence of motional-excited states.

Next, we compute single-molecule and two-molecule tweezer-bound states to determine the binding energies $\epsilon$ and interaction energy $U$ for microwave shielded molecules in experimentally realistic systems.
We consider an isotropic Gaussian tweezer potential $V = -D\exp\left(-\frac{2r_1^2}{w_0^2}\right)$,
where $D$ and $w_0$ are the tweezer depth and waist, and $r_1$ is the distance of the molecule to the center of the tweezer.
Single-molecule and two-molecule tweezer-bound states are computed using sinc-function discrete variable representation (sinc-DVR) \cite{colbert1992novel}, as described in detail in the Appendix.

Experimentally, many routes are accessible for forming tight, near-isotropic optical traps.
Red-detuned tweezers [see Fig.~\ref{fig:1}(a)] can be generated by acousto-optic modulators or spatial light modulators (SLM), projected from a high-numerical-aperture objective. An additional standing-wave lattice (not shown) to confine the axial direction results in a near-isotropic trap.
For even tighter waists, a blue-detuned, short-wavelength trap could be used, where molecules are repelled from high intensity regions, requiring the generation of an ``anti-tweezer'' potential. 
This could be done with an SLM such that low-intensity regions form potential minima for trapping~\cite{trisnadi2022design}. 
Then, a blue-detuned axial lattice confines the third dimension. 
Finally, we note that our proposal is extensible to three-dimensional red- and blue-detuned optical lattices as well.

The results of the bound state calculations are shown in Fig.~\ref{fig:dvr}.
For NaCs, KAg, and FrAg, we choose
realistic tweezer waists assuming either a blue-detuned or a red-detuned trap laser, given the respective species' strong electronic transitions~\cite{Cairncross2021, Stevenson2023, tomza, klos2022prospects}, and assume a diffraction-limited spot size is achievable ($w_0 = 0.61 \lambda/\mathrm{NA}$) with microscope numerical aperture of NA$\sim$0.6.
Solid lines show the binding energy, $\epsilon$, of the single-molecule ground state.
Dotted lines show $U-\epsilon$, the difference of the two-molecule and single-molecule ground state energies.
At the lowest depths where the single-molecule bound state has appeared, there are no two-body bound states [blue shaded region in Fig.~2(a)].
As $D$ rises to intermediate values, two-molecule bound states exist but $U-\epsilon < \epsilon$ [white shaded region in Fig.~2(a)].
For KAg and FrAg this region spans more than one decade in tweezer depth,
and within this region both $\epsilon$ and $U-\epsilon$ increase with depth,
leading to favorable conditions for deterministic single-particle loading.
For NaCs, this region is extremely narrow and does not feature a maximum in $U-\epsilon$.
At the highest depths, the binding energy exceeds the repulsion [red shaded region in Fig.~2(a)].
This sets a maximum $D$ at which there is an energetic cost to loading a second molecule.

\begin{figure}[t]
    \centering
    \includegraphics[width=0.475\textwidth]{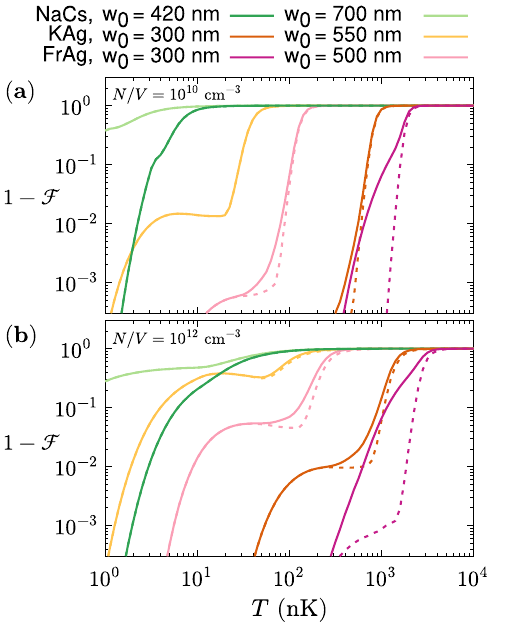}
\caption{
{\bf Deterministic loading of shielded molecules} for fixed densities $10^{10}$ and $10^{12}$\,cm$^{-3}$ in panel ({\bf a}) and ({\bf b}), respectively.  
Dashed lines show the infidelity $1-\mathcal{F}$ for single molecule preparation;
solid lines show $1-\mathcal{F}$ for preparing that single molecule in its motional ground state. 
Near perfect loading is possible for highly dipolar molecules and for small-waist tweezers.
Full dependence on temperature and phase-space density is shown in the Appendix.
\label{fig:infidelity}
}
\end{figure}

Next we discuss the probability to populate traps with single molecules [shown in Fig.~\ref{fig:infidelity}], with $\epsilon,U$ obtained from the bound state calculations.
At a fixed temperature, we pick the optimum depth, $D$, by maximizing the fidelity of single-molecule preparation with the constraint that $2\epsilon>U>\epsilon$.
This optimum trap depth essentially coincides with the maximum of $U-\epsilon$, seen in Fig.~\ref{fig:dvr}.
Infidelity is dominated by loading zero or two molecules [see Appendix].
For fixed reservoir gas densities of $10^{10}$\,cm$^{-3}$ and $10^{12}$\,cm$^{-3}$, the results are shown in Fig.~\ref{fig:infidelity} for varying temperatures as dashed lines.
For the highly polar KAg/FrAg in blue traps ($w_0=300$\,nm), the fidelity of single-molecule ground-state loading can be above 99\% at temperatures of $\sim100-1000$\,nK.
Many other alkali-metal-coinage-metal molecules have intermediate dipolar length~\cite{Smialkowski}, and are expected to show similar performance.
In larger tweezers of $w_0>500\,$nm, this region of good single-molecule loading falls toward lower absolute temperature, and the infidelities are comparatively worse (but still ${>}90\%$ across a broad $T$ range) -- confirming the intuition that smaller trap volumes (compared to $R_6$) are better for deterministic single-particle  loading.
For less polar NaCs in a $w_0=420$~nm tweezer, deterministic loading of ${>}90$\% can be realized with achievable temperatures of several nanokelvin,
but loading with large 700\,nm waists will not produce fidelities above $\sim 30-40$\% for even $T=1$\,nK.

In addition to deterministic single-particle loading, our scheme also enables population of the 3D motional ground state of the tweezer potential with high fidelity, as shown as the solid lines in Fig.~\ref{fig:infidelity}.
For KAg and FrAg in small tweezers, the fidelity for a single ground-state molecule exceeds 90\% at $\sim$400~nK and exceeds 99\% at $\sim$200~nK.
For NaCs, the ground-state loading curves are overlapped with the single-particle loading curves.

We anticipate several pathways to further improve the fidelity.
First, the interaction blockade can be enhanced 
by using a repulsive dipolar $1/r^3$ potential in a 2D reservoir trap with all molecules polarized in the axial direction~\cite{valtolina2020dipolar}, as opposed to compensated microwave-shielded $1/r^6$ interactions in a 3D trap as we have considered here.
Additionally, loading from a 2D reservoir that is nearly matched in density to the tweezer array may provide better initial phase-space overlap compared to loading from a 3D reservoir, potentially enabling higher fidelities.
Finally, careful tailoring of the tweezer turn-on ramp may improve fidelities beyond our equilibrium description, which hence sets a lower limit to the performance.

We have proposed a scheme for preparation of low-entropy tweezer arrays of polar molecules in the motional ground state.
Here, optimal trap depths are shallower than those used for direct laser cooling, leading to potentially larger array sizes.
Furthermore, our scheme is general for bosons and fermions.
Indeed,
with Pauli blocking of spin-polarized fermions, our technique is made more robust against multiparticle occupancies, though we do not rely on this.
Our proposal does not require quantum degeneracy: while degenerate Fermi gases and Bose-Einstein condensates of polar molecules have recently been produced~\cite{demarco2019degenerate, schindewolf2022evaporation, bigagli2024observation, shi2025bose}, quantum degenerate molecules still present a technical overhead, realized by only a handful of experiments.
Our scheme could also facilitate the preparation of atom arrays by Rydberg-dressing~\cite{Johnson2010} to realize an interaction blockade analogous to the one we discussed for polar molecules.
Our proposal could introduce for a new paradigm in loading large-scale arrays for quantum information science.

We thank Jacob Covey, Mohit Verma, and Dave DeMille for stimulating discussions.
We thank Kang-Kuen Ni for careful reading of the manuscript.
Z.Y.~acknowledges support from the David and Lucile Packard Foundation (grant 2024-77404), the AFOSR Young Investigator Program (grant FA9550-25-1-0360), and the Neubauer Family Assistant Professors Program. S.W.~acknowledges support from NSF (Award No.~2409747), an AFOSR (Award No.~FA9550-25-1-0048), and the Gordon and Betty Moore Foundation (Award No.~GBMF12340). T.K.~acknowledges NWO VIDI (Grant ID 10.61686/AKJWK33335).

\clearpage
\bibliography{ref}

\clearpage
\appendix
\section{Calculation of binding energies, interaction energies, and fidelity of single-molecule preparation}

We describe the two-molecule states in center of mass coordinates $\bm{R} = (\bm{r}_1 + \bm{r}_2)/2$ and $\bm{r} = \bm{r}_1-\bm{r}_2$.
This makes treating an interaction that depends only on $\bm{r}$ simple.
We write the isotropic Gaussian trapping potential as the Legendre expansion
\begin{align}
V =& -2D\exp\left(-\frac{2R^2+\frac{r^2}{2}}{w_0^2}\right) \nonumber \\
&\times \sum_{\mathrm{even}~\ell} (2\ell+1)~i_\ell\left(\frac{2rR}{w_0^2}\right)~P_\ell(\cos\theta),
\end{align}
where $\theta$ is the angle between $\bm{R}$ and $\bm{r}$,
$P_\ell(z)$ is a Legendre polynomial,
and $i_\ell(x)$ is a modified spherical Bessel function of the first kind, related to the modified Bessel function of the first kind as $i_\ell(x) = \sqrt{\pi/2x} I_{\ell+1/2}(x)$.
Note that this finite depth potential is not separable in center of mass and relative coordinates.

We expand the two-molecule bound state in a basis set of Clebsch-Gordan-coupled spherical harmonics 
\begin{align}
|l L J M \rangle = \sum_{m_l,M_L} |l m_l \rangle| L M_L \rangle\langle l m_l L M_L | J M \rangle 
\end{align}
where $\langle \bm{r} | l m_l \rangle  = Y_{l,m_l}(\hat{\bm{r}})$ and $\langle \bm{R} | L M_L \rangle  = Y_{L,M_L}(\hat{\bm{R}})$.
For an isotropic trap and interaction, $J$ and $M$ are good quantum numbers, and the calculation is completely independent of the value of $M$.
We truncate the basis as $l,L \le 6.$
For the matrix element of the tweezer potential we will need
\begin{align}
\langle l L J M | P_\ell | l' L' J M \rangle = (-1)^{l+l'+J} [l,l',L,L']^{1/2} \nonumber \\
\times \sixj{l}{l'}{\ell}{L'}{L}{J} 
\threejm{l}{0}{\ell}{0}{l'}{0} \threejm{L}{0}{\ell}{0}{L'}{0},
\end{align}
where $[l,l',\ldots] = (2l+1)(2l'+1)\ldots$, and the quantity in curly brackets is a Wigner $6j$ symbol.
The matrix element above vanishes if $\ell > l+l'$ or $\ell > L+L'$,
such that in a truncated basis set only a limited number of terms in the expansion of the tweezer potential contribute.

We treat the $r$ and $R$ radial degrees of freedom numerically using sinc-DVR \cite{colbert1992novel}.
We first treat these degrees of freedom separately to compute $20$ contractions, using as a reference potentials $V = -D \exp\left(-\frac{r^2}{2w_0^2}\right)$ and $V=- D\exp\left(-\frac{2R^2}{w_0^2}\right)$.
Then, we set up a basis set of direct products of radial contractions and Clebsch-Gordan-coupled spherical harmonics.
We compute the lowest bound states of the full Hamiltonian, including the coupling between the center of mass and relative coordinate, computed as explained above.
As an interaction we include a repulsive isotropic van der Waals interaction $c_6 r^{-6}$ to model collisional shielding \cite{karman2025double}.
The $c_6$ coefficient is chosen to lead to $R_6 = 3000~a_0$ for NaCs,
appropriate for double microwave shielding with compensated dipolar interactions as realized experimentally \cite{bigagli2024observation,zhang2025observation}.
For KAg we take $R_6 = 9000~a_0$, reflecting the three times larger dipolar length than for NaCs, following the universality of microwave shielding \cite{dutta2025universality}.
For FrAg we take $R_6 = 14\,000$~$a_0$, which is also obtained using universal scaling,
but reducing the $c_6$ coefficient by one order of magnitude to model shielding at reduced Rabi frequencies that are an order of magnitude larger\cite{karman2025double}.
This is done to keep the Rabi frequencies around one MHz,
rather than scaling universally with $E_6 \propto R_6^{-2}$ which could lead to Rabi frequencies comparable to light or Zeeman shifts, for example.

We compute the single-molecule energy levels, $E^{(1)}_i$, and two-molecule energy levels, $E^{(2)}_i$, as described above.
We relate the single-molecule energy levels to the binding energy as $E^{(1)}_i = - \epsilon_i$,
where the zero of energy is the potential energy in the reservoir.
That is, $\epsilon_i$ is the binding energy of the $i$th single-particle state and $\epsilon\equiv \epsilon_0$ is the ground state binding energy.
We write the two-molecule energy levels as $E_i^{(2)} = -2\epsilon_0+U_i$, to obtain the state-dependent repulsive interaction, $U_i$.
In practice, we obtain the single-particle energy levels as the energy minus half the ground state energy of two non-interacting molecules.
The energy levels of the non-interacting molecules are obtained using the same computational procedure to maximize a systematic cancellation of errors and to prevent that any difference in accuracy of the treatment of the single-molecule and two-molecule problem shows up as a spurious interaction.
Then, the population of any single-molecule state is given by
\begin{align}
\frac{p^{(1)}_i}{p^{(0)}} = z \exp\left(\frac{\epsilon_i}{k_BT}\right),
\end{align}
and the population of any two-molecule state is given by
\begin{align}
\frac{p^{(2)}_i}{p^{(0)}} = z^2 \exp\left(\frac{2\epsilon_0-U_i}{k_BT}\right).
\end{align}
The fidelity for single-molecule preparation, shown in Fig.~\ref{fig:infidelity},
is then computed as
\begin{align}
\mathcal{F}^\mathrm{sp} = \frac{1}{\mathcal{Q}} \sum_i z \exp\left(\frac{\epsilon_i}{k_BT}\right),
\end{align}
and for ground-state preparation
\begin{align}
\mathcal{F}^\mathrm{gs} = \frac{1}{\mathcal{Q}} z \exp\left(\frac{\epsilon_0}{k_BT}\right),
\end{align}
where
\begin{align}
\mathcal{Q}= 1+\sum_i z \exp\left(\frac{\epsilon_i}{k_BT}\right) +\sum_i z^2 \exp\left(\frac{2\epsilon_0-U_i}{k_BT}\right).
\end{align}
The fugacity $z$ is determined by $\mathrm{Li}_{3/2}(z) = \rho$, where $\mathrm{Li}_{3/2}$ is the poly-logarithmic function,
and $z\approx \rho$ for phase-space density much lower than unity \cite{huang1963statistical}.

We provide a simple estimate of the occurrence of higher-than-double occupation.
Assuming that the repulsion between molecules is pairwise additive,
this gives
\begin{align}
\frac{p^{(3)}}{p^{(2)}} = z \exp\left( \frac{\epsilon -2 U}{k_BT} \right).
\end{align}
This is smaller than $p^{(2)}/p^{(1)}$ by a factor $\exp(-\frac{U}{k_BT})$,
which means that in the regime $U>\epsilon\gg k_BT$ where double occupancy is suppressed,
higher occupancy can be safely neglected.

\begin{figure*}
    \centering
    \includegraphics[width=0.475\textwidth]{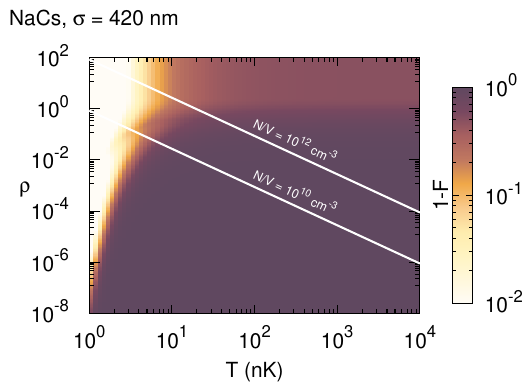}
    \includegraphics[width=0.475\textwidth]{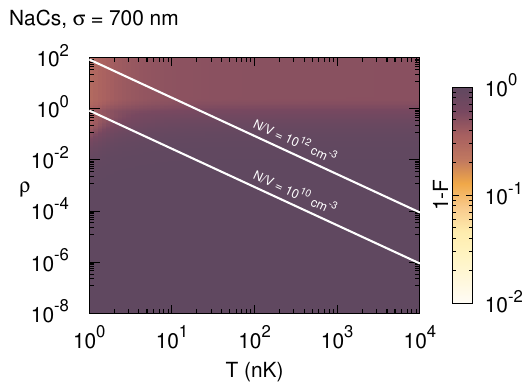}
    \includegraphics[width=0.475\textwidth]{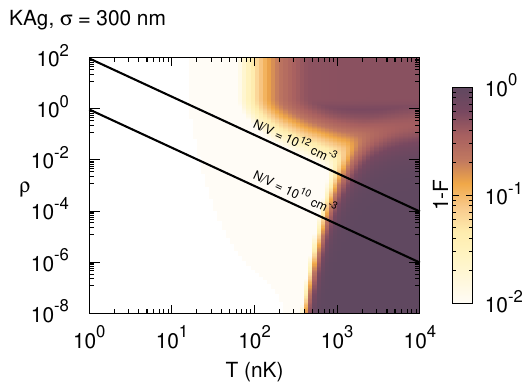}
    \includegraphics[width=0.475\textwidth]{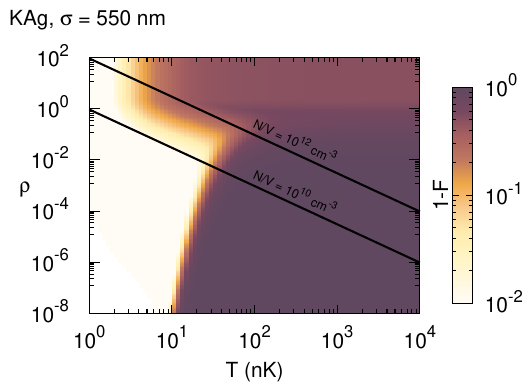}
    \includegraphics[width=0.475\textwidth]{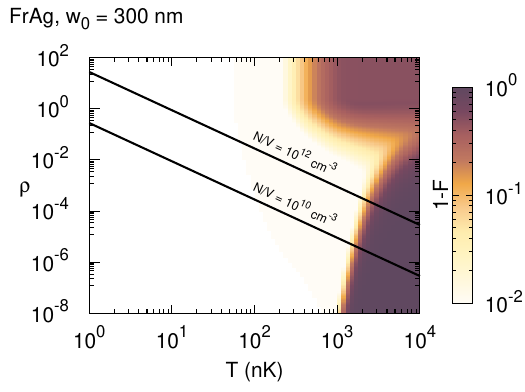}
    \includegraphics[width=0.475\textwidth]{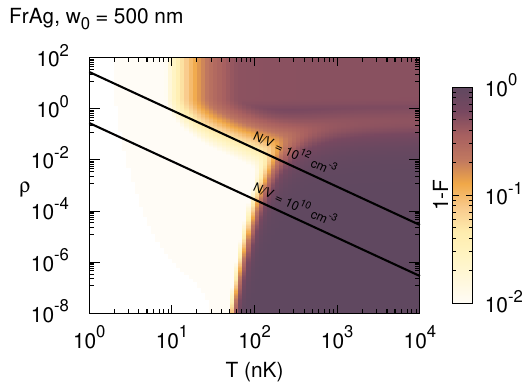}
\caption{
Infidelity of deterministic single molecule preparation as a function of phase space density and temperature.
Lines indicate the region of constant densities of $10^{10}$ and $10^{12}$~cm$^{-3}$, respectively, used in the main text Fig.~3.
}
    \label{fig:fid1_constrained}
\end{figure*}

\begin{figure*}
    \centering
    \includegraphics[width=0.475\textwidth]{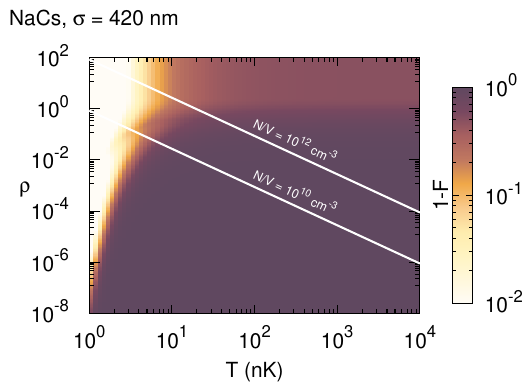}
    \includegraphics[width=0.475\textwidth]{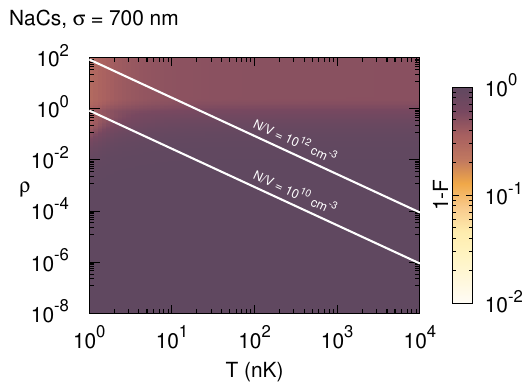}
    \includegraphics[width=0.475\textwidth]{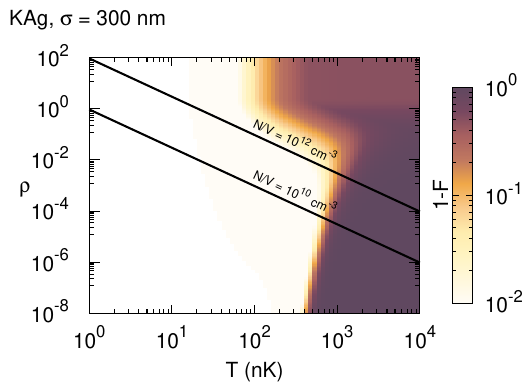}
    \includegraphics[width=0.475\textwidth]{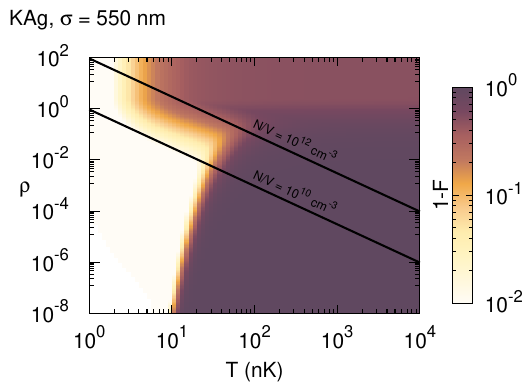}
    \includegraphics[width=0.475\textwidth]{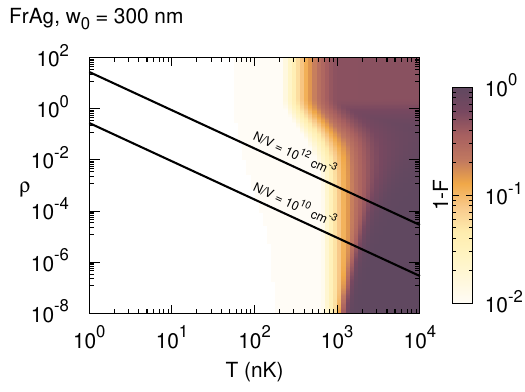}
    \includegraphics[width=0.475\textwidth]{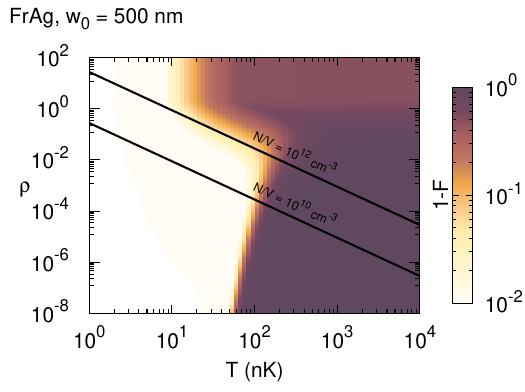} 
\caption{
Infidelity of single molecule preparation in the motional ground state as a function of phase space density and temperature.
Lines indicate the region of constant densities of $10^{10}$ and $10^{12}$~cm$^{-3}$, respectively, used in the main text Fig.~3.
}
    \label{fig:fidgs_constrained}
\end{figure*}

\section{Inelastic and elastic collision rates}

Here we give estimates on the collision rates between the reservoir molecules, between the reservoir and the tweezer-trapped molecules, and between two tweezer molecules, focusing on regimes identified as promising for low-entropy array loading.
First, we discuss the robustness of our proposed scheme to inelastic loss mechanisms, which could undermine our assumptions of equilibrium if losses far outpace typical loading rates.
Under conditions of optimal double microwave shielding, this is not the case.
Collisions with molecules from the bulk reservoir gas will occur on similar time scales for single tweezer-trapped molecules and molecules in the bulk gas,
as both are set by the density of the bulk gas.
For NaCs, both two- and three-body loss can be made negligible on a 10~s timescale with double microwave shielding at bulk densities as high as $10^{12}$\,cm$^{-3}$~\cite{yuan2025extreme}.
For the ultrapolar molecules, the two-body rates are predicted to be even slower, around $10^{-15}$ cm$^3$/s for KAg~\cite{karman2025double}.

Considering  double occupancies in tweezers, the intra-tweezer density could be far higher than the bulk density.
If the interparticle spacing is set by the harmonic oscillator length scale, $\sqrt{\hbar/2m\omega}$, where $\omega/2\pi=40\,$kHz is the trap frequency, the optimal depth of $D=h\times200$\,kHz trap for KAg with $w_0=300$\,nm would result in a density of maximally $\sim10^{16}$\,cm$^{-3}$ and therefore a $>0.1$s predicted lifetime.
The tweezer would not be constantly hosting two or more particles, so this lifetime should be taken as a lower bound, and is sufficiently long for loading from a reservoir gas.

The elastic collision rate sets the time scale for tweezer loading.
The cross section is estimated as $\sigma = 4\pi R_6^2$,
which yields $0.3$, $2.9$, and $7.2$ square micron for NaCs, KAg, and FrAg, respectively.
The rate coefficient $\beta = v\sigma$ where $v=\sqrt{\frac{2kT}{m}}$ is a typical velocity,
and the time scale for collisions is then related to the density as $\tau=1/n\beta$.
For NaCs at 1~nK temperature where loading can be effective, this leads to a 1 second timescale for collisions at $10^{10}$~cm$^{-3}$ density.
This is somewhat shorter than the $\sim$10~second lifetime of the gas.
At the higher density of $10^{12}$~cm$^{-3}$ that we consider, the timescale for elastic collisions is correspondingly reduced to 10~ms.
For KAg and FrAg, where both the cross section is larger and loading can be effective at 100~nK temperature, elastic collisions occur with 100~$\mu$s and 60~$\mu$s time constants even at $10^{12}$~cm$^{-3}$ density.
Therefore the approximation of thermal equilibrium from frequent collisions is justified in these cases.

\end{document}